# Parton Skewed Distributions in the Pion and Quark-Hadron Duality

ALEXANDER P. BAKULEV, RUSKO RUSKOV *

*Bogolyubov Laboratory of Theoretical Physics, JINR, 141980 Dubna, Russia*

and

KLAUS GOEKE, N. G. STEFANIS

*Institut für Theoretische Physik II,
Ruhr-Universität Bochum, D-44780 Bochum, Germany,*

Applying arguments based on the operator product expansion for a three-point correlator and relying on quark-hadron duality, we derive an expression for the skewed (nonforward) parton distribution in the pion in the case of a zero-skewedness parameter, $\mathcal{F}_{\zeta=0}^{\psi|\pi}(X;t)$. We expect that our result is relevant for moderately large momentum transfers $1 \lesssim t \lesssim 10$ GeV$^2$. In addition, we construct a purely phenomenological factorized model for the same quantity in close analogy to Radyushkin's model, originally proposed for skewed distributions of quarks in the nucleon. Though the quark-hadron duality approach supports theoretically the factorized model, the two models exhibit a different behavior in the parton momentum fraction $X$ at any fixed $t$. The relevant process to distinguish between the two options seems to be the WACS off the pion that measures (to leading $t/s$-order) the inverse moment $<X^{-1}>$ of the skewed distribution. Even after the inclusion of the first order kinematic $t/s$-corrections, the predictions for the cross section $\frac{d\sigma}{dt}(s,t)$ at c.m.s. scattering angles $\vartheta = 30°$ and $90°$ differ by factors 3.5–3.9 and 2.9–7.5, respectively, so that a discrimination appears possible.

## I. INTRODUCTORY REMARKS

Recently, Radyushkin [1] has argued that at moderately large momentum transfer $t = (1 - 10)$ GeV$^2$, hadronic form factors and wide-angle Compton scattering (WACS) amplitudes are dominated by a soft mechanism corresponding to an overlap of soft wave functions. This analysis was performed in terms of universal nonforward parton densities $\mathcal{F}(X;t)$, which accumulate the soft contribution in the WACS case. These densities are obtained in the $\zeta = 0$ limit of the nonforward parton distributions (NFPDs), $\mathcal{F}_\zeta(X;t)$, introduced in [2,3],[†] and represent the simplest hybrid distribution interpolating between usual parton densities $f_\psi(x)$ and hadronic form factors. A simple factorized model for $\mathcal{F}(X;t)$ in the proton was constructed in [1] using the well-known Glück–Reya–Vogt parameterization [7] for $f_{\psi|\text{proton}}(x)$ and assuming a Gaussian dependence on the transverse momentum $k_\perp$ of the effective two-body soft light-cone wave function $\Psi(x, k_\perp)$ of the proton.

---

*On leave of absence from the Institute for Nuclear Research and Nuclear Energy, 1784 Sofia, Bulgaria.
[†]The NFPDs are similar to but not coinciding with the off-forward parton distributions (OFPDs) introduced by Ji in [4,5]; cf. the discussion in [3,6].

In this paper, we perform a similar analysis relating to the case of the nonforward parton densities in the pion. We show that the factorized model can be approximately justified within the Operator Product Expansion (OPE) in conjunction with QCD sum rules [8].

As a first step towards a complete QCD sum-rule analysis, we explore in this paper the so-called local quark-hadron duality approximation (simply abbreviated as LD in the following) [8–12] that was successfully applied to estimate various nonperturbative characteristics, like hadron masses, leptonic widths, electromagnetic form factors of hadrons, etc. [10–17]. Note that our approach has the advantage of being simultaneously gauge and Lorentz invariant from the outset.

Within this framework, we are able to obtain a compact expression for the NFPDs, like $\mathcal{F}_{\zeta=0}(X;t)$, in the form of an overlap integral of the Drell–Yan–West-type [18,19], which involves an effective two-body soft pion wave function $\Psi(x, \vec{\mathbf{k}}_\perp)$, introduced earlier in [12]. We compare the LD expression for $\mathcal{F}_{\zeta=0}(X;t)$ with the corresponding factorized ansatz and comment on the reliability of the LD approximation.

Finally, an estimate of the leading contribution to WACS off the pion is made for moderately large scattering angles (in the c.m. frame). We also discuss possible corrections of order of $\mathcal{O}(t/s)$ *at leading twist* and show that new nonperturbative quantities should be introduced in addition to the skewed parton densities. These new quantities are particular $y$-moments of the same underlying double distributions $F(x, y; t)$.

The analysis of these moments, as well as an extension of our approach to the nucleon case, will be done elsewhere.

The paper is organized as follows. In Sect. II we establish our definitions of double distributions (DDs) for the pion case. We derive their symmetry properties and relations to the corresponding NFPDs. In Sect. III the operator product expansion approach for the skewed distributions in the pion is discussed in connection with local duality. The prediction for the distribution $\mathcal{F}_{\zeta=0}(X;t)$ following from this approach, is given in Sect. IV. The consistency of the LD result with the general sum rule for the pion form factor is demonstrated. Sect. V deals with the modeling of the skewed distributions, using a phenomenological parameterization for the valence quark distribution in the pion. We demonstrate that the WACS process off the pion is very sensitive to the model for $\mathcal{F}_{\zeta=0}(X;t)$, especially to its small $X$ behavior. In Sect. VI we discuss finite $\mathcal{O}(t/s)$ corrections to the handbag contribution of the skewed distributions. Finally, in Sect. VII we further discuss our results and draw our conclusions.

## II. DOUBLE DISTRIBUTIONS IN THE PION; DEFINITIONS AND SUM RULES

First, we define double distributions in the pion in close analogy to the nucleon case [2,3] in terms of a nonforward matrix element of a bilocal quark-antiquark operator on the light cone using, however, the most general decomposition (see, also [20])

$$\langle \pi(p') | \bar{\psi}(0)\gamma_\mu E(0, z; A)\psi(z) | \pi(p)\rangle|_{z^2=0} =$$

$$(p+p')_\mu \int_0^1 \int_0^1 \theta(x+y \leq 1) \left(e^{-ix(pz)-iy(rz)} F_{\psi|\pi}(x, y; t) - e^{ix(pz)-i\bar{y}(rz)} F_{\bar{\psi}|\pi}(x, y; t)\right) dx\, dy$$

$$+ (p-p')_\mu \int_0^1 \int_0^1 \theta(x+y \leq 1) \left(e^{-ix(pz)-iy(rz)} G_{\psi|\pi}(x, y; t) - e^{ix(pz)-i\bar{y}(rz)} G_{\bar{\psi}|\pi}(x, y; t)\right) dx\, dy \quad (1)$$

$$+ iz_\mu \int_0^1 \int_0^1 \theta(x+y \leq 1) \left(e^{-ix(pz)-iy(rz)} Z_{\psi|\pi}(x, y; t) - e^{ix(pz)-i\bar{y}(rz)} Z_{\bar{\psi}|\pi}(x, y; t)\right) dx\, dy,$$

where $r \equiv p - p'$ is the momentum transfer ($r^2 \equiv -t < 0$) and $E(0, z; A) = P \exp(ig \int_z^0 dx_\mu A_\mu(x))$ is the path-ordered gauge string factor in the fundamental representation ($A_\mu \equiv \sum_{a=1}^8 t_a A_\mu^a$).[‡] Eq.(1) corresponds to operator structures of leading and next-to-leading twists (twist 2 and 3, respectively).

Matrix elements of this type appear in the perturbative QCD analysis of deeply virtual Compton scattering (DVCS) processes [4,2]. One can actually prove in all orders of perturbation theory that the large-$Q^2$ asymptotics of the scattering amplitude can be represented in a factorized form with the short-distance part calculated perturbatively (see, e.g., [22,3]). The dynamics of large distances, which is mainly nonperturbative, is in turn accumulated in the matrix elements of the type given by Eq.(1).

---

[‡]Note that to leading-twist accuracy, the definition of the matrix element in Eq.(1) is independent of the choice of the contour connecting the quark fields $\bar{\psi}(0), \psi(z)$ (cf., e.g., [21]).

The $z_\mu$-term in Eq.(1) is of higher twist-3, that is, it will produce a power-suppressed contribution ($\sim 1/Q^2$) in the DVCS amplitude. Another two-body twist-3 part is produced by the axial-vector matrix element

$$\langle \pi(p')|\bar{\psi}(0)\gamma_\mu\gamma_5 E(0,z;A)\psi(z)|\pi(p)\rangle|_{z^2=0} =$$
$$p_\alpha r_\beta z_\sigma \epsilon_{\mu\alpha\beta\sigma} \int_0^1 \int_0^1 \theta(x+y\leq 1)\left(e^{-ix(pz)-iy(rz)}A_{\psi|\pi}(x,y;t) - e^{ix(pz)-i\bar{y}(rz)}A_{\bar{\psi}|\pi}(x,y;t)\right) dx\, dy \quad . \quad (2)$$

Because we neglect in this paper power-suppressed corrections, our main focus will be on the $F$-, $G$-distributions.

The parameterization of the nonforward matrix element in terms of DDs is natural and can be established, at least in perturbation theory, at any order of $\alpha_s$ [2,3]. Just as in the case of deep-inelastic scattering, DDs in Eq.(1) have a parton interpretation: $F_{\psi|\pi}(x,y;t)$ is the amplitude to find an active quark in the pion with momentum fractions $x$ and $y$ of the initial (hadron) momentum $p$ and the momentum transfer $r$, respectively.

Due to the conservation of the local vector current, the DDs $G_{\psi|\pi}(x,y;t)$ obey a sum rule. Indeed, for $z=0$ only the first term in Eq.(1) should survive ($r^2 \neq 0$) and this implies

$$\int_0^1 \int_0^1 \theta(x+y\leq 1)\left(G_{\psi|\pi}(x,y;t) - G_{\bar{\psi}|\pi}(x,y;t)\right) dx\, dy = 0. \quad (3)$$

Moreover, there are stronger sum rules:

$$\int_0^1 \int_0^1 \theta(x+y\leq 1)\, G_{\psi,\bar{\psi}|\pi}(x,y;t)\, dx\, dy = 0. \quad (4)$$

Indeed, using TP-invariance and complex conjugation of the matrix elements in Eqs.(1),(2), one can easily show that the DDs introduced above are real-valued functions and should obey the symmetry relations

$$\begin{aligned} F_{\psi|\pi}(x,1-x-y;t) &= F_{\psi|\pi}(x,y;t) \\ G_{\psi|\pi}(x,1-x-y;t) &= -G_{\psi|\pi}(x,y;t) \\ Z_{\psi|\pi}(x,1-x-y;t) &= Z_{\psi|\pi}(x,y;t) \\ A_{\psi|\pi}(x,1-x-y;t) &= A_{\psi|\pi}(x,y;t) \; , \end{aligned} \quad (5)$$

which are a generalization of the "Münich symmetry" relations [23] in the case of the pion. Hence, the sum rules encoded in Eq.(4) are an obvious consequence of that symmetry for the DD $G_{\psi|\pi}(x,y;t)$.

Within the generalized Bjorken limit for DVCS, we have the relations $Q^2, pq' \gg t, m_\pi^2$ and $r_\parallel = \zeta p$ [4,2], where $\zeta \equiv Q^2/2pq'$ coincides with the Bjorken variable. Considering $\zeta$ as an external parameter, one can introduce the NFPDs [2,3] $\mathcal{F}_\zeta(X;t)$, with $X = x + \zeta y$ being the total fraction of the momentum of the active parton, to read

$$\mathcal{F}_\zeta^{\psi|\pi}(X;t) = \theta(X\geq \zeta)\int_0^{\bar{X}/\bar{\zeta}} F_{\psi|\pi}(X-y\zeta,y;t)\, dy + \theta(X\leq \zeta)\int_0^{X/\zeta} F_{\psi|\pi}(X-y\zeta,y;t)\, dy \; . \quad (6)$$

Analogously, one can also define the "forward invisible" NFPDs $\mathcal{G}_\zeta^{\psi|\pi}(X;t)$, $\mathcal{Z}_\zeta^{\psi|\pi}(X;t)$, $\mathcal{A}_\zeta^{\psi|\pi}(X;t)$.

Then the leading contribution of the handbag diagrams in Fig. 1 can be obtained in the form

$$T_{\mu\nu}(p,q,q') = \frac{1}{2}\sum_\psi e_\psi^2 \left(-g_{\mu\nu} + \frac{1}{p\cdot q'}(p_\mu q'_\nu + p_\nu q'_\mu)\right) \quad (7)$$

$$\times \int_0^1 dX \left[\frac{1}{X-\zeta+i\epsilon} + \frac{1}{X-i\epsilon}\right]\left[(\zeta-2)\left(\mathcal{F}_\zeta^\psi(X;t) + \mathcal{F}_\zeta^{\bar{\psi}}(X;t)\right) - \zeta\left(\mathcal{G}_\zeta^\psi(X;t) + \mathcal{G}_\zeta^{\bar{\psi}}(X;t)\right)\right] \; .$$

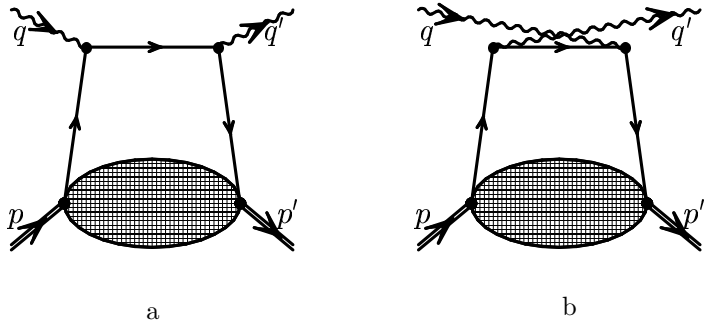

FIG. 1. Diagrams contributing to the DVCS amplitude. The blobs at the bottom correspond to DDs.

Another process, described with the same handbag diagrams, is the wide-angle Compton scattering (WACS), recently reexamined in papers [1,24] for the proton case. Now the initial photon is also real ($Q^2 = 0$), but $t \equiv r^2$ is large enough to ensure the light-cone dominance (see [1] for a discussion of other contributions with subleading $\mathcal{O}(t/s)$ behavior). The contribution of large distances in this case will be described by the same skewed distributions at $\zeta = 0$, notably

$$\mathcal{F}^{\psi|\pi}_{\zeta=0}(X;t) \equiv \mathcal{F}^{\psi|\pi}(X;t). \tag{8}$$

Of course, in this case one should use Eq.(7) with some care because finite $t$-corrections to the hard part may become important [15,1]. These $t$-corrections, which are within the leading-twist approximation, are analogous to the target-mass corrections in DIS which have led to the Nachtmann–Georgi–Politzer $\xi$-scaling [25,26].

As it was shown in refs. [2,3], DDs play a key role in describing those processes in which nonforward matrix elements are involved. In fact, many properties of the skewed distributions, like polynomiality [27,28], symmetry properties [23,28], etc., can be simply established using the integral representations of the type of Eq.(6).

However, DDs seem to have a more complicated structure. In fact, the question of possible singularities of the DDs is still open. A pure perturbative analysis seems to yield DDs without singularities [3]. However, DDs are by definition nonperturbative objects and one may expect such singularities to appear in the real world. In the work of [28], $t$-channel meson-exchange contributions were considered and found to produce $\delta$-function type singularities. In the case of pion DDs a dynamical mechanism was found [20], based on the effective chiral model which follows from the instanton vacuum of QCD [29,30].

Independently of whether or not these singularities really appear in the DDs, it is clear that the skewed distributions, which are certain integrals of the DDs, are more smoothly-behaved functions. Thus, they seem to be more appropriate for modeling.

### III. OPE APPROACH TO THE SKEWED DISTRIBUTIONS

In this and the subsequent sections, we are going to study the skewed distribution in the pion, $\mathcal{F}^{\psi|\pi}_{\zeta}(X;t)$, within an approach which is based on QCD sum rules [8], with particular emphasis being placed on the case $\zeta = 0$. As in the proton case [4,5], one can derive a sum rule connecting the charge pion form factor with a certain integral of the NFPD (the zeroth moment), namely,

$$F_\pi(t) = \sum_\psi e_\psi \int_0^1 \left[ \mathcal{F}^{\psi|\pi}_\zeta(X;t) - \mathcal{F}^{\bar\psi|\pi}_\zeta(X;t) \right] dX , \tag{9}$$

where $e_\psi$ is the electric charge of the active "$\psi$"-quark (see also [31]). Indeed, Eq.(9) follows immediately from Eq.(1) after taking the limit $z = 0$ and incorporating the sum rule for the DD, termed $G$ in Eq.(3). It should be emphasized, however, that for fixed $\zeta \neq 0$, the physical domains for the $t$-variable on the left and the right hand side of Eq.(9) are different. In fact, for the DVCS process, the limit $t \to 0$ is unreachable [32], i.e.,

$$t \geq t_{\min} = \frac{\zeta^2 m_\pi^2}{1-\zeta}. \tag{10}$$

On the other hand, in the forward limit ($\zeta = 0$, $t = 0$), a reduction formula holds, (see [4,2]), and we have

$$\mathcal{F}^{\psi|\pi}_{\zeta=0}(X; t=0) = f_{\psi|\pi}(X) . \tag{11}$$

Both, the form factor, as well as the parton distribution in the middle region of $X$, were thoroughly investigated within the QCD sum-rule approach. In fact, for the pion form factor $F_\pi(t)$ it was shown that in the region of momentum transfers, $t \geq 1$ GeV$^2$, the so-called Feynman mechanism [33] is capable to reproduce the experimental data [13,9] without recourse to the hard part.

In this paper we shall adopt a similar philosophy and use to derive the skewed distribution in the pion the concept of local quark-hadron duality [8]- [17] that was successfully applied to the calculation of various nonperturbative characteristics, like hadron masses, leptonic widths, electromagnetic form factors, etc. (for some recent applications, we refer to [15,12,16,17]).

As usual in the QCD sum rule approach, let us consider the three-point amplitude

$$R^u_{\alpha\mu\beta}(p, p'; z) = i^2 \int d^4x \int d^4y \, e^{-ipx} e^{ip'y} \left\langle 0 \left| T \left\{ j^{5+}_\alpha(x) \, \bar{u}(0) \gamma_\mu E(0,z;A) u(z) \, j^5_\beta(y) \right\} \right| 0 \right\rangle , \tag{12}$$

where $j^5_\alpha(x) = \bar{d}(x) \gamma_5 \gamma_\alpha u(x)$ is the axial current with a nonzero projection on the pion state, so that

$$\langle 0 | j^5_\alpha(0) | \pi^+(p) \rangle = i f_\pi p_\alpha , \quad f_\pi \simeq 133 \text{ MeV}, \tag{13}$$

and $z$ denotes a light-like coordinate ($z^2 = 0$).

The correlator defined by Eq.(12) will be considered in the Euclidean region for $p, p', r \equiv p - p'$. It gives contributions to different invariant form factors with tensor structures proportional to $p_\alpha(p+p')_\mu p'_\beta$, $p_\alpha r_\mu p'_\beta$, $g_{\alpha\beta} r_\mu$, etc. For the $\zeta = 0$ case we shall project on a light-like direction via $n_\alpha n_\beta n_\mu$, where $n^2 = 0$, $p^+ \equiv (np)$, and $r^+ \equiv (nr) = 0$. The advantage of this projector is that it projects out the leading structure in the infinite momentum frame (IMF), where $p^+ \to \infty$, with $r_\perp$ fixed. This structure is also most directly related to the one analyzed before in [9,34] by employing the QCD sum-rule method to calculate the pion form factor.

Picking out the invariant amplitude of the leading structure, we have

$$R_u(p^2, p'^2, t; z) = \frac{1}{\pi^2} \int_0^\infty \int_0^\infty \frac{\rho^{\text{phys}}_u(s, s', t; z)}{(s-p^2)(s'-p'^2)} \, ds \, ds' + \cdots , \tag{14}$$

where the ellipsis denotes polynomials in $p^2, p'^2$. The perturbative contribution to $R_u(p^2, p'^2, t; z)$ (which is the leading term of an OPE expansion in the deeply Euclidean region of the momentum invariants) can be written in the same form as (14) with the obvious change $\rho^{\text{phys}}_u \to \rho^{\text{pert}}_u$. Due to asymptotic freedom, for large $s$ and $s'$, $\rho^{\text{phys}}_u(s, s') \sim \rho^{\text{pert}}_u(s, s')$. However, for small $s, s'$, the two spectral densities differ drastically from each other. Indeed, $\rho^{\text{phys}}_u$ contains the pion double $\delta$-function term

$$\rho^\pi_u(s, s', t, z) = 2\pi^2 f^2_\pi \, \delta^{(+)}(s - m^2_\pi) \, \delta^{(+)}(s' - m^2_\pi)$$
$$\times \int_0^1 \left( e^{-iX(pz)} \mathcal{F}^{u|\pi}_{\zeta=0}(X;t) - e^{iX(pz)} \mathcal{F}^{\bar{u}|\pi}_{\zeta=0}(X;t) \right) \, dX , \tag{15}$$

whereas, in contrast, $\rho^{\text{pert}}_u(s, s')$ is a smooth function for any *finite* order of perturbation theory.

Under the proviso of the local duality (LD) assumption, one has that $\rho^\pi_u(s, s')$ is dual to $\rho^{\text{pert}}_u(s, s')$ in an appropriate duality interval, so that

$$\frac{1}{\pi^2} \int_0^{s_0} \int_0^{s_0} \rho^{\text{pert}}_u(s, s') \, ds \, ds' = \frac{1}{\pi^2} \int_0^{s_0} \int_0^{s_0} \rho^\pi_u(s, s') \, ds \, ds'. \tag{16}$$

Here the duality interval $s_0$ corresponds to the effective threshold of the higher excited states and the "continuum" in the channels with the quantum numbers of the axial current.

The LD relation (16) is very natural within the QCD sum-rule approach [8]. In fact, the effective threshold $s_0$ is fixed by the ratio of the nonperturbative power corrections (the condensate contributions) relative to the (leading) perturbative term in the OPE for the correlator (cf. Eq.(12)). In what follows, we shall use the value

$$s_0 = s_0^{\text{LD}} = 4\pi^2 f_\pi^2 \ , \tag{17}$$

which follows from the LD prescription for the correlator of two axial currents:

$$\Pi_{\alpha\beta}(p) = i \int e^{-ipx} \left\langle 0 \left| T \left\{ j_\alpha^{5+}(x) j_\beta^5(0) \right\} \right| 0 \right\rangle d^4x \ . \tag{18}$$

For the experimental value of $f_\pi$, we have $s_0^{\text{LD}} \cong 0.67$ GeV$^2$. This value is very close to the standard one, $s_0^{\text{SR}} \approx 0.7$ GeV$^2$, that has been extracted from the direct QCD sum-rule approach for the 2-point correlator, Eq.(18), in the classical work of Ref. [8], in determining the pion decay constant $f_\pi$.

The same duality interval was also obtained in the QCD sum-rule analysis of Refs. [13,9] of the charged pion form factor at moderate momentum transfers $t \approx 1-3$ GeV$^2$. However, it was observed in [35] that for higher values of $t$, the relative contribution of the condensate (power) corrections increase, ensuing an increase of the extracted parameter $s_0$ as well. This situation corresponds to the so-called infrared regime [36,15], i.e., to the kinematical regime, where one of the quarks carries most of the momentum of the initial hadron (current). In this regime the underlying OPE series becomes badly convergent and should be resumed in some way, e.g., by introducing nonlocal condensates $\langle \bar{q}(0)q(x)\rangle$, $\langle G(0)G(x)\rangle$, $\langle \bar{q}(0)G(x)q(y)\rangle$, etc. [37,38]. Adopting a reasonable model for the nonlocal condensates, it was shown [35] that the form factor $F_\pi(t)$, extracted from such an improved approach, can describe the data up to $t \sim 10$ GeV$^2$ in compliance with previous rough estimations [39,11], according to which the asymptotically leading hard-scattering contribution starts to become important beyond $t \gtrsim 10$ GeV$^2$. As discussed more fully in [35], the corresponding threshold $s_0^{(3 \text{ GeV}^2 \leq t \leq 10 \text{ GeV}^2)}$ in such a type of QCD sum-rule analysis was found to have approximately the standard value, quoted above.

## IV. LOCAL DUALITY PREDICTIONS FOR THE SKEWED DISTRIBUTIONS IN THE PION

The one-loop contribution to the double spectral density can most easily be calculated using light-cone variables in a frame where the initial momentum $p$ has no transverse components, i.e.,

$$p = \left\{ p^+, \ p^- = \frac{s}{p^+}, \ \vec{0}_\perp \right\} \tag{19}$$

and the momentum transfer has no "plus" component (cf. [12]):

$$r = \left\{ 0, \ r^- = \frac{s}{p^+} - \frac{s' + \vec{r}_\perp^{\ 2}}{p^+}, \ \vec{r}_\perp \right\} \ , \quad k = \left\{ Xp^+, \ k^-, \ \vec{k}_\perp \right\} \ . \tag{20}$$

Here, $s, s'$ are the invariant masses in the channels with pion quantum numbers, $X$ is the total fraction of the longitudinal momentum carried by the quark entering the composite vertex, and $\vec{k}_\perp$ is its transverse momentum.

Applying the Cutcosky rules (cf. Fig. 2a), one can obtain for the double spectral density of the "$R_u$"-correlator (recall Eq.(14)) in leading order of $\alpha_s$

$$\rho_u^{\text{pert}}(s, s', t; z) = \frac{3}{\pi} \int_0^1 e^{-iX(pz)} dX \int \delta\left(s - \frac{\vec{k}_\perp^{\ 2}}{X\bar{X}}\right) \delta\left(s' - \frac{\left(\vec{k}_\perp - \vec{r}_\perp \bar{X}\right)^2}{X\bar{X}}\right) d^2\vec{k}_\perp \ . \tag{21}$$

For the contribution at hand, the spectral constraint $0 < X < 1$ reflects the positivity of the energy of the struck and spectator quark, respectively. In fact, this constraint has a more general nature, discussed, for instance, in [3,5].

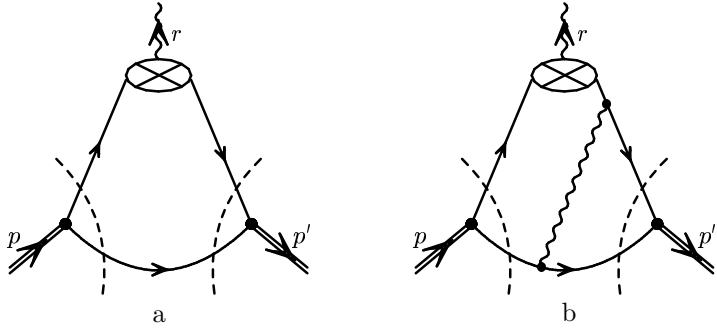

FIG. 2. Typical Cutcosky cuts (dashed lines) for perturbative diagrams in the OPE for the 3-point correlator of two hadron currents, involving a composite operator (crossed oval at the top). The left and right graphs correspond to the $O(1)$ and $O(\alpha_\text{s})$ contributions, respectively.

Substituting Eq.(21) into the LD relation, provided by Eq.(16) one can extract (by taking the Fourier transform) the corresponding skewed distribution

$$\mathcal{F}^{\text{LD}}_{\zeta=0,u|\pi}(X;t) = \frac{3}{4f_\pi^2\pi^3} \int \Theta\left(s_0 - \frac{\vec{\mathbf{k}}_\perp^{\,2}}{X\bar{X}}\right) \Theta\left(s_0 - \frac{\left(\vec{\mathbf{k}}_\perp - \vec{\mathbf{r}}_\perp\bar{X}\right)^2}{X\bar{X}}\right) d^2\vec{\mathbf{k}}_\perp . \qquad (22)$$

Analogously, one can deduce that, in leading order, the following relations hold

$$\begin{aligned}\mathcal{F}^{\text{LD}}_{\zeta=0,\bar{d}|\pi}(X;t) &= \mathcal{F}^{\text{LD}}_{\zeta=0,u|\pi}(X;t) \\ \mathcal{F}^{\text{LD}}_{\zeta=0,\bar{u}|\pi}(X;t) &= \mathcal{F}^{\text{LD}}_{\zeta=0,d|\pi}(X;t) = 0.\end{aligned} \qquad (23)$$

Eq.(22) can be recast in the form of an overlap integral of effective two-body soft light-cone wave functions of the incoming and outgoing pion:

$$\mathcal{F}^{\text{LD}}_{\zeta=0,u|\pi}(X;t) = \int \psi^{\text{LD}}\left(X,\vec{\mathbf{k}}_\perp\right) \psi^{\text{LD}}\left(X,\vec{\mathbf{k}}_\perp - \bar{X}\vec{\mathbf{r}}_\perp\right) \frac{d^2\vec{\mathbf{k}}_\perp}{16\pi^3} . \qquad (24)$$

The explicit form of the effective wave function $\psi^{\text{LD}}(x,\vec{\mathbf{k}}_\perp)$ can be obtained, in particular, from the evaluation of the 2-point correlator, Eq.(18), to read [12]

$$\psi^{\text{LD}}_{\text{eff}}(X,\vec{\mathbf{k}}_\perp) = \frac{2\sqrt{6}}{f_\pi} \Theta\left(s_0 - \frac{\vec{\mathbf{k}}_\perp^{\,2}}{X\bar{X}}\right) . \qquad (25)$$

Thus, the LD prescription reproduces an expression of the Drell–Yan–West type [18,19]. In a recent paper [24], the classical Drell–Yan–West formula was used as a starting point for developing a phenomenological model for the form factors and skewed distribution of the proton.

Carrying out the integration over the transverse momentum $\vec{\mathbf{k}}_\perp$, we obtain from Eq.(24)

$$\mathcal{F}^{\text{LD}}_{u|\pi}(X;t) = \frac{3}{\pi^3 f_\pi^2} \Theta\left(X - \frac{t}{4s_0+t}\right) s_0 X\bar{X} \left[\arccos\sqrt{\frac{\bar{X}t}{X4s_0}} - \sqrt{\frac{\bar{X}t}{X4s_0}\left(1 - \frac{\bar{X}t}{X4s_0}\right)}\right] . \qquad (26)$$

Note that the $\Theta$- function in Eq.(26) is due to the corresponding abrupt behavior of the effective pion wave function (25) dictated by local duality at leading order. It ensures that the corresponding $X^{-1}$-moment of $\mathcal{F}^{\text{LD}}_{u|\pi}(X;t)$, which enters the leading handbag expression for the WACS, is indeed finite (cf. Eq.(7)). We expect that at next-to-leading (NLO) order in $\alpha_s$ (cf. Fig. 2b) of the LD-approach this property will remain intact. In fact, such a property is required by the general considerations of the factorization theorem [22,3].

For small momentum transfers $t$, the form of the quark-hadron duality will change (see the discussion bellow). However, in a more realistic model for the effective wave function and/or skewed distribution, this property should be preserved as well.

Taking the zero-order $X$-moment of $\mathcal{F}_{u|\pi}^{\text{LD}}(X;t)$ and incorporating the relations (23), one can reproduce to leading-order accuracy the general sum rule Eq.(9) (at $\zeta = 0$)

$$F_\pi^{\text{LD}}(t) = \int_0^1 \mathcal{F}_{u|\pi}^{\text{LD}}(X;t) \, dX \;, \tag{27}$$

where

$$F_\pi^{\text{LD}}(t) = 1 - \frac{1 + 6s_0/t}{(1 + 4s_0/t)^{3/2}} \;. \tag{28}$$

Actually, the explicit LD prediction for the pion form factor was obtained earlier in [9,12].

Thus, the concept of local duality provides us with nontrivial dynamical information about the form of the pion wave function/skewed distribution, as well as about the behavior of the pion form factor at moderately large momentum transfers $t \gtrsim 0.6$ GeV$^2$. As we shall see in the next section, this form-factor prediction seems to be supported by the existing experimental data in this momentum region. Moreover, due to the Ward identity, connecting the 3-point function Eq.(12) and the 2-point function, Eq.(18), the property $F_\pi(t=0) = 1$ is also preserved.[§]

On the other hand, it is well known that in the region of small momentum transfer $t$, the quark-hadron duality is more complicated [40,9,41,15]. Thus, one should not overestimate the accuracy of Eq.(27) in the region of $t \leq s_0$. In fact, the derivative of Eq.(27) at zero momentum transfer is infinite. The reason is that in the kinematical region $t \ll |p^2|, |p'^2|$, one has to include additional terms in the OPE corresponding to the situation in which the current with small momentum transfer is placed at a large distance. This leads to the notion of bilocal power corrections [40,9]. Taking them into account extends the validity of the theoretical QCD sum-rule prediction for the pion electromagnetic form factor to the whole momentum transfer region $t = (0 - 3)$ GeV$^2$, providing, in particular, the correct value of the pion charge radius $\frac{dF}{dt}|_{t=0}$ [9].

It is important to note that the same reasoning can be applied to a more complicated object, namely, the skewed distribution $\mathcal{F}_{u|\pi}(X;t)$. In fact, the $t = 0$ limit of Eq.(24) gives

$$\mathcal{F}_{u|\pi}^{\text{LD}}(X;t=0) = f_{u|\pi}^{\text{LD}}(X) = 6X\bar{X}. \tag{29}$$

The parton distribution $f_{u|\pi}^{\text{LD}}(X)$ is normalized to unity and hence respects the same Ward identity, mentioned above. Moreover, it turns out that its form coincides with the asymptotic (leading-twist) distribution amplitude of the pion $\varphi_\pi^{\text{as}}(x)$. Thus, as a consequence, the (naive) quark-hadron duality procedure fails to reproduce a reasonable valence parton density in the pion. Actually, to reproduce the small $t$-behavior of the skewed distribution $\mathcal{F}_{u|\pi}^{\text{LD}}(X;t)$ within a QCD sum rule approach is a rather complicated problem. Both, the notions of bilocal power corrections of leading twist [15], as well as non-local condensates [37,38] should be introduced. We shall address this interesting problem in a forthcoming publication.

However, for large momentum transfers $t \geq 1$ GeV$^2$, in analogy to the form factor calculation, one may expect that the LD result, given by Eq.(26), should work.

## V. MODELING THE SKEWED DISTRIBUTIONS

In order to get some experience of how reliable the LD strategy is, and to estimate how large the deviations in the region of small momentum transfers $t$ may be, we shall derive in this section a factorized model for the skewed parton distribution in the pion at $\zeta = 0$.

---

[§]This property was shown to be fulfilled in a complete QCD sum-rule analysis for the pion form factor, performed in [34].

We shall argue that the factorized ansatz for $\mathcal{F}^{u|\pi}_{\zeta=0}(X;t)$, following the approach of Ref. [1], is of the form

$$\mathcal{F}^{u|\pi;\text{Factorized}}_{\zeta=0}(X;t) = f_{u|\pi}(X)\, e^{-t\bar{X}/2\Lambda^2 X} \ , \tag{30}$$

which makes it apparent that it automatically satisfies the general "reduction relation" (11), [4,2,3]. Note that $f_{u|\pi}(X)$ is the valence $u$-quark distribution in the pion.

The specific functional dependence of the exponential in (30) on $t$- and $X$- can be formally justified within the OPE approach. In fact, if one replaces into the overlap formula (24) the abrupt LD wave function, $\psi^{\text{LD}}(X,\vec{\mathbf{k}}_\perp)$, given by Eq.(25), by the popular Gaussian ansatz, proposed in [42,43],

$$\psi^{\text{Gaussian}}(X,\vec{\mathbf{k}}_\perp) = \Phi(X)\, e^{-\vec{\mathbf{k}}_\perp^2/2\Lambda^2 X\bar{X}} \ , \tag{31}$$

one also arrives at such an exponential dependence. Another hint at an exponential dependence can be traced back to the double Borel transform of the 3-point correlator (12) when employing the OPE. In fact, the $t$-dependence of the perturbative term (first diagram on the rhs of Fig. 3) and that of the term involving a (vector) nonlocal quark condensate (inserted into the bottom line of the second diagram on the rhs of the same figure) is described by one and the same function, namely,

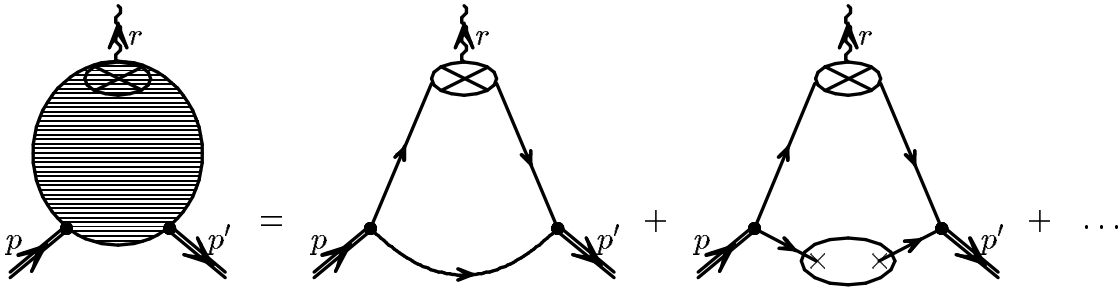

FIG. 3. Operator product expansion of the 3-point correlator of two pion currents involving a composite operator.

$$\Phi(t,X) \equiv e^{-t\bar{X}/(M_1^2 + M_2^2)X} \ . \tag{32}$$

Here $M_1^2$ and $M_2^2$ are Borel parameters, corresponding to $p^2$ and $p'^2$, respectively, and $X$ is the momentum fraction, flowing through the upper lines. Other terms of the nonlocal OPE – not displayed in the figure – are more complicated, but numerically their $t$-dependence is similar.

A Gaussian distribution form for the transverse momentum should not be surprising. It is in line with the Borel transformation technique, as it was recently shown in [44], where effective Gaussian wave functions were derived.

The dimensional parameter $\Lambda$ in (30) should be determined in correspondence with some averaged values of the Borel parameters $\langle M_{1,2}^2\rangle$ for which the underlying sum rule is saturated. Thus, one may expect that

$$\Lambda^2 \approx \langle M_{1,2}^2 \rangle \approx 2\cdot 0.7\ \text{GeV}^2 = 1.4\ \text{GeV}^2 \ , \tag{33}$$

where $0.7\ \text{GeV}^2$ is the characteristic scale for the two-point pion correlator and its 3-point counterpart is twice larger [35].

On the other hand, $\Lambda$ can be determined on a purely phenomenological ground by adopting the Glück–Reya–Schienbein (GRSch) parameterization for the valence quark distribution in the pion [47]:

$$x f_{u|\pi}(x) = 0.5645\, x^{0.504}(1 + 0.153\sqrt{x})(1-x)^{0.349} \ , \tag{34}$$

which refers to a rather low normalization point $\mu^2_{\text{LO}} = 0.26\ \text{GeV}^2$. According to our considerations, we shall use in our modeling procedure a parton distribution evolved to a more appropriate scale: $\mu^2 \simeq 1\ \text{GeV}^2$. Taking into account the leading order LD estimates for $\mathcal{F}^{\psi|\pi}_{\zeta=0}(X;t)$ (recall Eq.(22)), one is tempted to assume that the valence $\bar{d}$ distribution is $f_{\bar{d}|\pi}(x) \simeq f_{u|\pi}(x)$ and that the sea quark contributions are practically negligible. (This assumption seems indeed to be supported by experiment [48,47]).

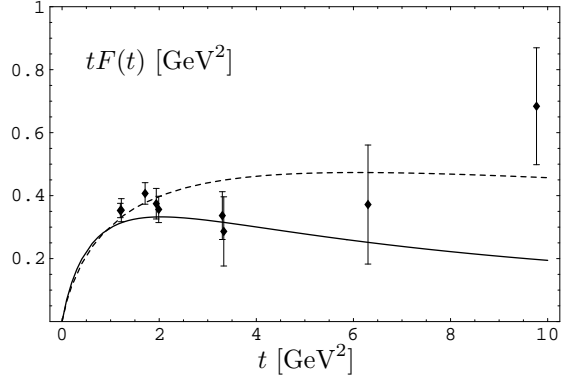

FIG. 4. Predictions for $tF_\pi(t)$ based on the LD model, Eq.(28), (solid line) and the factorized GRSch-ansatz, Eqs.(34)–(35) (dashed line). The experimental data are taken from [45]– [46].

To fix the parameter $\Lambda$, we use sum-rule relation (9), related to the electromagnetic pion form factor, and a properly weighted sum of quark and antiquark DDs. Taking into account (23) and $e_u - e_d = 1$, one can write

$$F_\pi^{\text{Factorized}}(t) \simeq \int_0^1 dX\, \mathcal{F}_{\zeta=0}^{u|\pi;\text{Factorized}}(X;t) = \int_0^1 dX\, f_{u|\pi}(X) \exp\left(\frac{-t\bar{X}}{2\Lambda^2 X}\right) , \qquad (35)$$

which makes the dependence on the parameter $\Lambda$ explicit. The best agreement between our factorized (and OPE-inspired) model (35) and the experimental data [45,46] in the region of intermediate momentum transfer $1 \text{ GeV}^2 \leq t \leq 10 \text{ GeV}^2$, is realized for $\Lambda_0^2 \approx 1.7 \text{ GeV}^2$ (see Fig. 4) which is not far away from the QCD sum-rule inspired value $\Lambda^2 = 1.4 \text{ GeV}^2$, given above.

Now, we are able to compare the two models in more detail. The results are presented in Fig. 5 for three different values of the momentum transfer: $t = 1, 3, 10 \text{ GeV}^2$. One observes from this figure a quite different behavior over the momentum fraction $X$ (at fixed momentum transfer $t$) for the two displayed models, though the general tendency is that for higher $t$ the distributions tend to shift their weight towards the upper end-point of the interval $0 < X < 1$. For the sake of completeness we also present in the same figure, a factorized model based on the naive LD distribution function, Eq.(29). As one may expect, this option gives a somewhat "smoother" behavior relative to the LD curve.

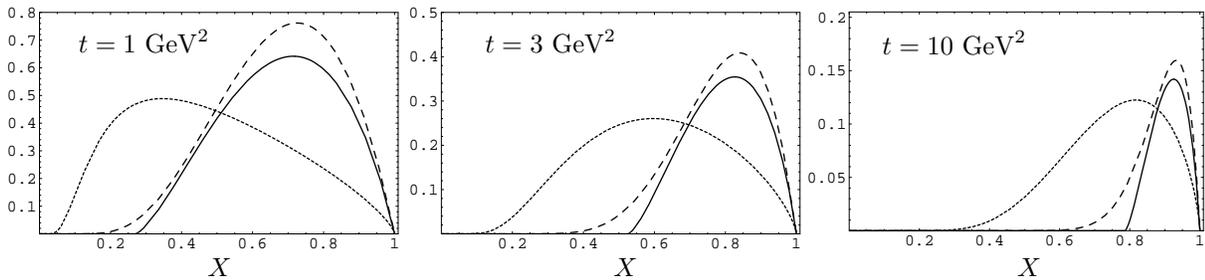

FIG. 5. Predictions for the skewed distributions in the pion with $\zeta = 0$, obtained with local duality (solid lines) and the factorized ansatz in conjunction with the GRSch parameterization (dotted lines), and the (naive) LD parton distributions, Eq.(29), (dashed lines).

By construction, the zero-order $X$-moment following from the factorized ansatz for $\mathcal{F}_{\zeta=0}^{\psi|\pi}(X;t)$, (cf. Eq.(35)), fits the data for the pion form factor rather well. On the other hand, also the LD prediction for

the pion form factor (Eq.(28)) complies with the data quite well. Thus, in order to distinguish between the two models, one should look for other physical observables, which are more sensitive to the form of the distribution.

In principle, the form of the true distribution at given $t$ can be reconstructed having recourse to higher moments: $<X^N>$ ($N>0$). In addition, the small $X$-behavior of $\mathcal{F}_{\zeta=0}^{\psi|\pi}(X;t)$ is sensitive to the inverse moments $<X^{-N}>$. Remarkably, a possibility to measure the $<X^{-1}>$ is offered by the WACS process in the pion case. Indeed, as we have mentioned above, the leading part of the handbag contribution to the WACS amplitude is proportional to the integral

$$<X^{-1}>\equiv R_{-1}^u(t) = \int_0^1 \mathcal{F}_{\zeta=0}^{u|\pi}(X;t)\frac{dX}{X} , \qquad (36)$$

as it can be seen from Eq.(7) at $\zeta=0$.

In the case of unpolarized initial photons, the differential cross section reads

$$\frac{d\sigma}{dt}(s;t) = \frac{4\pi\alpha^2}{s^2}\left(e_u^2+e_d^2\right)^2\left(R_{-1}^u(t)+R_{-1}^{\bar{u}}(t)\right)^2 , \qquad (37)$$

where we are still neglecting finite $t$-corrections. As we see from this equation, the inverse moment $X^{-1}$, or equivalently $R_{-1}^u(t)$, enters squared in the cross section formula. Thus, one may expect an improved sensitivity of the WACS process to different models of the skewed distribution in the pion (as well as also in other hadrons).

In Fig. 6 we show the differential cross sections of WACS off a pion as functions of the squared center-of-mass energy $s$ for different values of the scattering angle $\vartheta$ which is hold fixed. One infers from this figure that the two models presented above yield cross sections which differ from each other by approximately one order of magnitude.

We remark that estimates for the pion Compton scattering were also presented on phenomenological grounds in [49].

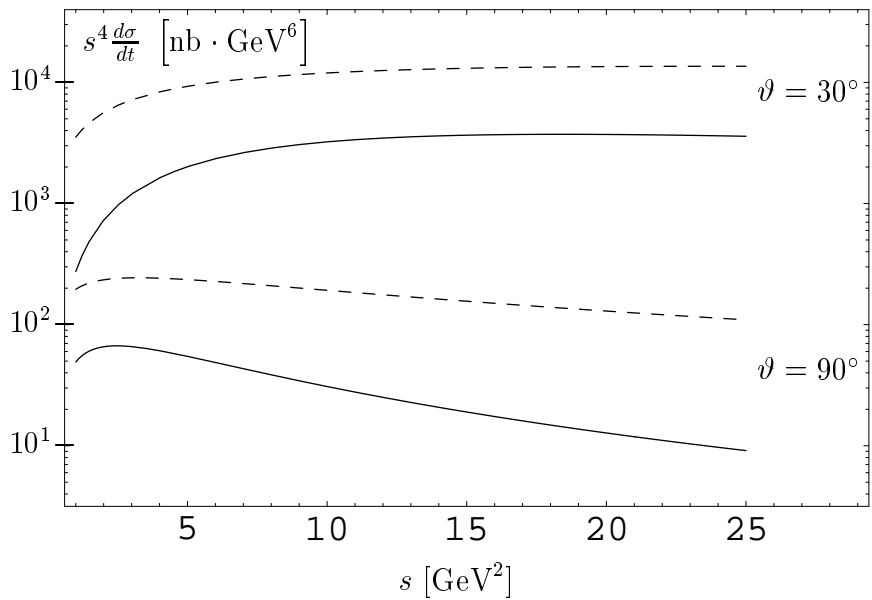

FIG. 6. Leading order predictions for the differential cross section of WACS off a pion, plotted against the squared center-of-mass energy $s$ for different values of the (fixed) scattering angle $\vartheta$ (in the center-of-mass reference system). The solid line shows the result derived from local duality; the dashed line that following from the GRSch-based model.

As we have mentioned in the Introduction, in order to ensure light-cone dominance, one should consider a Compton process at momentum transfers $t \gtrsim 1$ GeV$^2$. Thus, the $t/s$ ratio may not be small. We recall in this context that $t = \sin^2(\vartheta/2)s$ in the center-of-mass reference system, we are using, and $\vartheta$ is the c.m.s. scattering angle.

In such a situation, to make a more reliable comparison with the experimental data, one should actually take into account finite $t/s$ corrections. This task will be undertaken in the next section, where we consider possible corrections of order of $\mathcal{O}(t/s)$ at the leading-twist level.

## VI. FINITE $\mathcal{O}(t/s)$ CORRECTIONS WITHIN THE HANDBAG CONTRIBUTION

In order to derive the $\mathcal{O}(t/s)$ corrections at the leading-twist level, one can start from the contribution of the handbag diagram in the coordinate representation (see, [15]). Taking into account the parameterizations of the leading twist non-forward matrix elements through DDs ($F^\psi(x,y;t)$, $G^\psi(x,y;t)$, etc.), via Eq.(1), and performing a Fourier transformation, one arrives at the following expression for the Compton scattering amplitude

$$T_{\mu\nu}(p,q',r) = i \sum_{\psi=u,d} e_\psi^2 \int_0^1 dx \int_0^1 dy\, \theta(1 > x+y)\, S_{\mu\beta\nu\rho} \left\{ (p+p')_\rho \left( F^\psi(x,y;t) + F^{\bar\psi}(x,y;t) \right) + \right.$$

$$\left. + (p-p')_\rho \left( G^\psi(x,y;t) + G^{\bar\psi}(x,y;t) \right) \right\} (S_\beta(q_s) - S_\beta(q_u)) + \text{``A-term''} + \text{``Z-term''}, \quad (38)$$

where $S_{\mu\beta\nu\rho} \equiv g_{\mu\beta}g_{\nu\rho} - g_{\mu\nu}g_{\beta\rho} + g_{\mu\rho}g_{\nu\beta}$ and $S_\beta(q_s)$, $S_\beta(q_u)$ are, respectively, the hard quark propagators for the $s$ and $u$ channel handbag diagrams:

$$S_\beta(q) = \frac{q_\beta}{q^2}, \quad q_s = q + xp + yr, \quad q_u = q - xp + \bar y r. \quad (39)$$

For real Compton scattering, i.e., $q^2 = 0$, the denominators of the quark propagators can be written in the form

$$q_s^2 = xs - r^2 y(1-x-y) - x\bar x\, m_\pi^2 \quad \text{and} \quad q_u^2 = xu - r^2 y(1-x-y) - x\bar x\, m_\pi^2 \quad (40)$$

and respect the "Münich symmetry", Eq.(5).

In principle, it would be legitimate to retain in the calculation (see also [15]) $\mathcal{O}(m_\pi^2)$-, $\mathcal{O}(r^2)$-terms as well, in analogy to target-mass effects in DIS that have lead to $\xi$-scaling [25,26]. Here, in the pion case, the $m_\pi^2$-terms are not significant and can be neglected compared to $s$, $t$, and $u$.

Suppose for a moment that we neglect the $t$-corrections to the hard quark propagator. Then, we immediately reproduce the leading handbag contribution $T_{\mu\nu}(p,q',r)$, Eq.(7), with the leading tensor structure $-g_{\mu\nu} + \frac{1}{p \cdot q'}(p_\mu q'_\nu + p_\nu q'_\mu)$ depending on the two Sudakov 4-vectors $p,q'$. The appearance of the non-forward (skewed) distributions $\mathcal{F}_\zeta^{\psi|\pi}(X;t)$, $\mathcal{G}_\zeta^{\psi|\pi}(X;t)$, defined through the DDs $F_{\psi|\pi}(x,y;t)$, $G_{\psi|\pi}(x,y;t)$, in Eq.(6), is in accordance with the fact that in the *formal* $t \to 0$ limit, the denominators of the quark propagators depend on the combination $X = x + \zeta y$ only, the latter being the total momentum fraction of the active parton. For $\zeta = 0$, we have

$$\mathcal{F}^{\psi|\pi}(X;t) = \int_0^{\bar X} dy\, F_{\psi|\pi}(X,y;t), \quad \text{etc.} \quad (41)$$

In general, when $t \neq 0$, the scattering amplitude $T_{\mu\nu}(p,q',r)$ should depend on three independent 4-vectors which may be chosen to be $p,q',r_\perp$, or $p,q',r$, etc., and will also include non-leading tensor structures.** The denominators will also acquire an additional nontrivial dependence on the $x$- and, $y$-fractions.

It is in practice more convenient to consider directly the $(t/s)$-corrections to the cross section of WACS. For the case of unpolarized initial photons, and summing over the polarizations of the final one, we obtain[††]

---

**In fact $|r_\perp| \sim \sqrt{t}$.
[††] The $SU(2)$ symmetry relations (cf. Eq.(23)) have been taken into account.

$$\frac{d\sigma}{dt}(s;t) = \frac{4\pi\alpha^2}{s^2}\left(e_u^2 + e_d^2\right)^2 \left\{ \left(R_{-1}^u(t) + R_{-1}^{\bar{u}}(t)\right)^2 - \right.$$
$$- \frac{t}{s}\left[\left(R_{-1}^u(t) + R_{-1}^{\bar{u}}(t)\right)^2 - \left(R_{-1}^u(t) + R_{-1}^{\bar{u}}(t)\right)\left(R_1^u(t) + R_1^{\bar{u}}(t)\right)\right] +$$
$$\left. + \mathcal{O}\left(\frac{t^2}{s^2}\right)\right\}, \qquad (42)$$

where $R_1^u(t)$ is the corresponding first $X$-moment

$$< X^1 > \equiv R_1^u(t) = \int_0^1 dX\, X\, \mathcal{F}_{\zeta=0}^{u|\pi}(X;t)\,. \qquad (43)$$

The first term in Eq.(42) reproduces the leading result for the cross section, i.e., Eq.(37). It is worth remarking that the leading $t/s$-corrections are expressed only through the moments of the skewed parton distribution $\mathcal{F}_{\zeta=0}^{u|\pi}(X;t)$. In contrast, the NLO $t^2/s^2$-corrections include all distributions introduced in Section II.

Let us now introduce the $y$-moments of the DDs

$$\mathcal{F}_{\psi|\pi}^{(k)}(X;t) \equiv \int_0^{\bar{X}} dy\, y^k\, F_{\psi|\pi}(X,y;t) \qquad (44)$$

and analogously for the other DDs, namely, $G_{\psi|\pi}$, $A_{\psi|\pi}$, $Z_{\psi|\pi}$. Then the skewed distribution $\mathcal{F}^{\psi|\pi}(X;t)$ becomes simply the zeroth-order $y$-moment of the corresponding DD, $F_{\psi|\pi}(X,y;t)$. Moreover, one can show that the $t^2/s^2$-corrections can be expressed by means of the $X$-moments of these new distributions. Obviously, the higher $y$-moments, Eq.(44), cannot be expressed in terms of the simplest skewed distribution $\mathcal{F}^{\psi|\pi}(X;t)$. Thus, one is forced to introduce new skewed distributions, $\mathcal{F}_{\psi|\pi}^{(k\neq 0)}(X;t)$, though these distributions refer to the same DD. The generalization of Eq.(44) to non-zero skewedness, $\zeta \neq 0$, is straightforward but we shall skip it here for the sake of brevity.

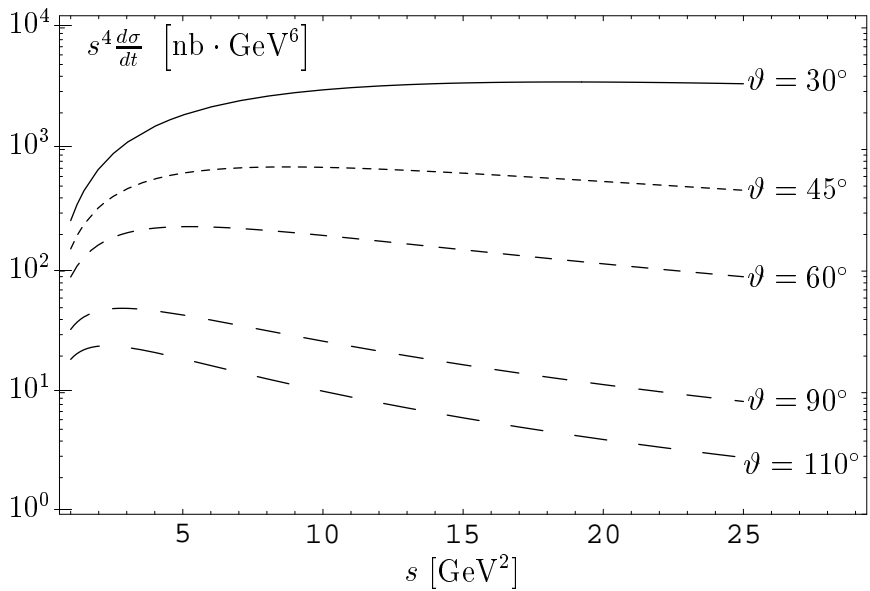

FIG. 7. Local duality (LD) predictions for the differential cross section of WACS off a pion with the first order kinematic $t/s$-corrections included as a function of the squared center-of-mass energy $s$ for selected values of the fixed scattering angle (in center-of-mass reference system) $\vartheta$.

The evaluation of the different $y$-moments is a separate task which we shall not address in this paper. Thus, we restrict our analysis here to the first order $t/s$-correction to the cross section, Eq.(42). The evaluation of the moments $R^u_{-1}(t)$, $R^u_1(t)$ is straightforward by virtue of Eqs.(26), (30).

Fig. 7 shows the plot of our LD predictions for the differential cross section of WACS off a pion as a function of the squared center-of-mass energy $s$ for different values of the fixed scattering angle (in the center-of-mass reference system) $\vartheta$. We expect that the prediction is reliable for $t \gtrsim 1$ GeV$^2$, as it was explained above. Notice here that the condition $t \gtrsim 1$ GeV$^2$ for the kinematics with fixed scattering angle $\vartheta$ transforms into the condition $s \gtrsim 1/\sin^2(\vartheta/2)$ GeV$^2$. For the minimal (30°) and maximal (90°) values of the scattering angle $\vartheta$ this means $s \gtrsim 15$ GeV$^2$ and $s \gtrsim 2$ GeV$^2$, respectively.

As it turns out, the relative magnitude of the $t/s$-correction to the LD-predicted skewed distribution $\mathcal{F}^{\text{LD}}_{u|\pi}(X;t)$ is $\approx 4\%$ for $\vartheta = 30°$ and rises to 10–28% for $\vartheta = 90°$. On the other hand, in the case of the factorized ansatz, $\mathcal{F}^{\text{Factorized}}_{u|\pi}(X;t)$, the corrections are stronger (5% and 25–42%, respectively).

Still, even after taking into account the $t/s$-corrections, the predictions for the cross section in the LD case, evaluated at $\vartheta = 30°$ (90°) is 3.5–3.9 (2.9–7.5) times smaller than with the corresponding factorized ansatz. Hence, forthcoming experiments at TJNAF [49] may be able to discriminate between these two models. The most dramatic difference of the two models appears, however, in the kinematical region $\vartheta = 90°$ and for $s \geq 10$ GeV$^2$.

## VII. DISCUSSION AND CONCLUSIONS

Let us start our discussion with some comments concerning the energy-momentum sum rule for the pion parton distributions, which we have rewritten in the form:

$$\int_0^1 dX\, X\, \left( \mathcal{F}^{u|\pi}(X;t) + \mathcal{F}^{\bar{d}|\pi}(X;t) + \mathcal{F}^{\bar{u}|\pi}(X;t) + \mathcal{F}^{d|\pi}(X;t) + \mathcal{F}^{g|\pi}(X;t) \right) |_{t=0} = 1 \;. \tag{45}$$

Here we have neglected the sea quark contribution of other flavors and $\mathcal{F}^{g|\pi}(X;t)$ is the $\zeta = 0$ version of the gluon skewed distribution in the pion that can be defined in an analogous way to the quark case (cf., e.g., [2]).

With the leading order LD result (meaning zeroth order in $\alpha_s$), i.e., Eqs.(26), (23), taken in the $t = 0$ limit, one observes that the sum rule, Eq.(45) is already saturated with the quark contribution alone; the sea and gluon contributions being exactly zero in this approximation.

In this way we demonstrate the consistency of the LD approach with the energy-momentum sum rule, Eq.(45). It is interesting to mention that the energy-momentum sum rule is saturated by the quark contribution alone also within the effective chiral model of Ref. [20]. Actually, their approach is based on a low-energy effective action derived from the instanton vacuum (see, [29,30] and references therein) and states that the gluon distribution should be parametrically small, $\sim (\bar{\rho}/\bar{R})^4$, where $\bar{\rho} \approx \frac{1}{600}$ MeV$^{-1}$ is the average instanton size and $\bar{R}$ is the average distance between instantons.‡‡

We would like to emphasize once more that the LD approach described in Section III is, strictly speaking, not applicable to the region of small momentum transfers $t \lesssim 0.6$ GeV$^2$ and should therefore be modified. For instance, using our factorized model for $\mathcal{F}^{\psi|\pi}(X;t)$, Eq.(30), which is based on the GRSch-parameterization of the quark distributions in the pion, one finds (at $\mu^2 \simeq 1$ GeV$^2$) for the quark contribution a value of $\sim 0.54$ to the energy-momentum sum rule with some room for the gluons left over. Note also that the contribution of the sea amounts only to $\sim 7\%$ of the total result.

In the region of applicability of our LD formula Eq.(26), $t \gtrsim 1$ GeV$^2$, we expect that the skewed distributions of the sea and the gluons should be suppressed. Indeed, within the LD approach they will both appear first at the level of $\alpha_s$-corrections.

Another point worth to be discussed concerns the estimation of the intrinsic transverse momentum of the pion. With the explicit form of the effective two-body wave function $\psi^{\text{LD}}_{\text{eff}}(X, \vec{k}_\perp)$, given by Eq.(25), one immediately obtains

---

‡‡The quark skewed distribution in the pion has been studied in the instanton vacuum, but in a somewhat different approach, also in Ref. [50].

$$\langle \vec{k}_\perp^2 \rangle_\pi^{\text{LD}} = \frac{s_0^{\text{LD}}}{10} \approx (260 \text{ MeV})^2 \;. \tag{46}$$

On the other hand, employing the factorized ansatz, Eq.(30), one gets for the valence quark distribution

$$\langle \vec{k}_\perp^2 \rangle_\pi^{\text{Factorized}} = \Lambda^2 \int_0^1 X\bar{X} f_{u|\pi}(X)\, dX \approx (320 \text{ MeV})^2 \;, \tag{47}$$

which is in moderate agreement with the LD estimate. In both cases a reasonable value of the intrinsic average transverse momentum of the pion is obtained. This value turns out to be somewhat smaller than that proposed by Kroll and collaborators [43]. The reason for this discrepancy may be traced back to the fact that the LD two-body wave function, Eq.(25), is an effective one and therefore includes not only the lowest particle-number (quark-antiquark) Fock state, but also an infinite tower of Fock states with additional soft gluons (cf. the discussion given in [12,44] and also [51]).

As a result, our $\psi_{\text{eff}}^{\text{LD}}(X, \vec{k}_\perp)$ can account for the full normalization condition

$$\int \frac{dX\, d^2\vec{k}_\perp}{16\pi^3} \left| \psi_\pi\left(X, \vec{k}_\perp\right)\right|^2 = 1 \;, \tag{48}$$

whereas the two-body wave function of Ref. [43] contributes approximately 1/4 of it.

In conclusion, we have provided evidence that the local duality approach for the skewed distribution $\mathcal{F}_{\zeta=0}^{u|\pi}(X;t)$ in the pion seems to support both, theoretically and phenomenologically, the so-called factorized model for the same quantity (proposed in [1] for the case of the proton). However, the two presented models show distinct behaviors in the longitudinal momentum fraction $X$ for any fixed momentum transfer $t$.

Furthermore, we have shown in this paper that measuring the Wide Angle Compton Scattering off the pion will provide us with a very sensitive tool to test the form of the skewed distribution and, in particular, to discriminate between the two models, proposed in this paper, and against others. We emphasize that our analysis was performed by taking into account the finite $t/s$-corrections to the cross section of WACS. Thus, we expect that our results may be relevant for an experimental check in the energy region of $s \simeq 2 \div 15$ GeV$^2$ and a scattering angle $\vartheta \simeq 30° \div 90°$ in the c.m.s. which seems accessible to the TJNAF machine [49].

In a forthcoming publication we plan to generalize the LD approach to the case of a non-zero skewedness parameter $\zeta$.

## Acknowledgments

This work was supported in part by the RFFI grant N 00-02-16696, by the Heisenberg–Landau Program, and by the COSY Forschungsprojekt Jülich/Goeke. We are grateful to A. V. Radyushkin who inspired this work and to S. V. Mikhailov, M. Polyakov, and C. Weiss for fruitful discussions. Two of us (A. B. and R. R.) are thankful to Prof. K. Goeke and his group for their warm hospitality at Bochum University, where part of this work was done.